\documentstyle[preprint,aps]{revtex}

\begin{document}

\draft


\title{New Black Hole Solutions in Brans-Dicke Theory of Gravity}

\author{Hongsu Kim\footnote{e-mail : hongsu@vega.kyungpook.ac.kr}}

\address{Department of Astronomy and Atmospheric Sciences \\
Kyungpook National University, Taegu, 702-701, KOREA}

\date{October, 1998}

\maketitle

\begin{abstract}
Existence check of non-trivial, stationary axisymmetric black hole solutions
in Brans-Dicke theory of gravity in different direction from those of Penrose,
Thorne and Dykla, and Hawking is performed. Namely, working directly with the
known explicit spacetime solutions in Brans-Dicke theory, it is found that
non-trivial Kerr-Newman-type black hole solutions {\it different} from general
relativistic solutions could occur for the generic Brans-Dicke parameter values 
$-5/2\leq \omega <-3/2$. Finally, issues like whether these new black holes carry
scalar hair and  can really arise in nature and if they can, what the associated physical
implications would be are discussed carefully.
\end{abstract}


\narrowtext

{\bf I. Introduction}
\\
 Of all the alternative theories of classical gravity to Einstein's general
relativity, perhaps the Brans-Dicke (BD) theory [1] is the most studied and
hence the best-known. This theory can be thought of as a minimal extension
of general relativity designed to properly accomodate both Mach's principle
[2] and Dirac's large number hypothesis [2]. Namely, the theory employs the
viewpoint in which the Newton's constant $G$ is allowed to vary with space
and time and can be written in terms of a scalar (``BD scalar'') field as
$G = 1/ \Phi $. In this work, we are interested in the existence of
exact solutions to the BD field equations that can describe rotating, charged
black hole spacetimes and their detailed structure. And if there are, we would
like to know whether they are non-trivial ones different from general
relativistic black hole solutions. As is well-known, even in Einstein's
general relativity, to find the exact solutions to the highly non-linear
Einstein field equations is a formidable task. For this reason, algorithms
generating exact, new solutions from the known solutions of simpler situations
have been actively looked for and actually quite a few were found. 
In BD theory of gravity, the field equations are even more complex and thus
it is natural to seek similar algorithms generating exact solutions from the
already known simpler solutions either of the BD theory or of the conventional
Einstein gravity. To the best of our knowledge, methods thus far discovered
along this line includes those of Janis et al., Buchdahl, McIntosh, Tupper,
Tiwari and Nayak, and Singh and Rai [3]. In particular, Tiwari and Nayak [3] proposed
an algorithm that allows us to generate stationary, axisymmetric solutions in
vacuum BD theory from the known Kerr solution [6] in vacuum Einstein theory and 
later on Singh and Rai [3] generalized this method to the one that generates 
stationary, axisymmetric, charged solutions in BD-Maxwell theory from the
known Kerr-Newman (KN) solution [6] in Einstein-Maxwell theory. Thus in the present work,
we shall take, as the Kerr-Newman-type solutions in BD-Maxwell theory (henceforth
``BDKN'' solutions), the ones constructed by Singh and Rai to explore if it can
descibe non-trivial black hole spacetimes {\it different} from those described by the
standard KN solution in Einstein-Maxwell theory. 
\\
{\bf II. Non-trivial BDKN black hole solutions}
\\
We begin by briefly reviewing
the algorithm proposed first by Tiwari and Nayak and generalized later by Singh 
and Rai. Consider the BD-Maxwell theory described by the action 
\begin{eqnarray}
S = \int d^4x \sqrt{g}\left[{1\over 16\pi}\left(\Phi R - \omega {{\nabla_{\alpha}\Phi
\nabla^{\alpha}\Phi }\over \Phi}\right) - {1\over 4}F_{\alpha \beta}F^{\alpha \beta}\right]
\end{eqnarray}
where $\Phi $ is the BD scalar field and $\omega $ is the generic parameter of the
theory. Extremizing this action then with respect to the metric $g_{\mu \nu}$, the
BD scalar field $\Phi $, and the Maxwell gauge field $A_{\mu}$ (with the field strength
$F_{\mu \nu}=\nabla_{\mu}A_{\nu}-\nabla_{\nu}A_{\mu}$) yields the classical field
equations given respectively by 
\begin{eqnarray}
G_{\mu \nu} &=& R_{\mu \nu} - {1\over 2}g_{\mu \nu}R = {8\pi \over \Phi}T^{M}_{\mu \nu}
+ 8\pi T^{BD}_{\mu \nu}, \nonumber \\
{\rm where} \\
T^{M}_{\mu \nu} &=& F_{\mu \alpha}F_{\nu}^{\alpha} - {1\over 4}g_{\mu \nu}F_{\alpha \beta}
F^{\alpha \beta}, \nonumber \\
T^{BD}_{\mu \nu} &=& {1\over 8\pi}\left[{\omega \over \Phi^2}(\nabla_{\mu}\Phi \nabla_{\nu}\Phi
- {1\over 2}g_{\mu \nu}\nabla_{\alpha}\Phi \nabla^{\alpha}\Phi) + {1\over \Phi}(\nabla_{\mu}
\nabla_{\nu}\Phi - g_{\mu \nu}\nabla_{\alpha}\nabla^{\alpha}\Phi)\right] \nonumber \\
{\rm and} \nonumber \\
\nabla_{\alpha}\nabla^{\alpha}\Phi &=& {8\pi \over (2\omega + 3)}T^{M\lambda}_{\lambda} = 0,
~~~~\nabla_{\mu}F^{\mu \nu} = 0, ~~~~\nabla_{\mu}\tilde{F}^{\mu \nu} = 0 \nonumber  
\end{eqnarray} 
with the last equation being the Bianchi identity and $\tilde{F}_{\mu \nu}={1\over 2}
\epsilon_{\mu \nu}^{\alpha \beta}F_{\alpha \beta}$. And the Einstein-Maxwell theory is
the $\omega \rightarrow \infty$ limit of this BD-Maxwell theory. Note that in the action and hence in the
classical field equations, there are no {\it direct} interactions between the BD scalar field
$\Phi$ and the ordinary matter, i.e., the Maxwell gauge field $A_{\mu}$. Indeed this is the 
essential feature of the BD scalar field $\Phi$ that distinguishes it from ``dilaton'' fields
in other scalar-tensor theories such as Kaluza-Klein theories or low-energy effective string
theories where the dilaton-matter couplings generically occur as a result of dimensional reduction.
(Here we would like to stress that we shall work in the context of original BD theory format not
some conformal transformation of it.)
As a matter of fact, it is the original spirit [1] of BD theory of gravity in which the BD scalar
field $\Phi$ is prescribed to remain strictly massless by forbidding its direct interaction
with matter fields. Now the algorithm of Tiwari and Nayak, and Singh and Rai goes as follows.
Let the metric for a stationary, axisymmetric, charged solution to Einstein-Maxwell field
equations take the form
\begin{eqnarray}
ds^2 = - e^{2U_{E}}(dt + W_{E}d\phi)^2 + e^{2(k_{E}-U_{E})}[(dx^1)^2 + (dx^2)^2] +
h^2_{E}e^{-2U_{E}}d\phi^2 
\end{eqnarray}
while the metric for a stationary, axisymmetric, charged solution to BD-Maxwell field 
equations be
\begin{eqnarray}
ds^2 = - e^{2U_{BD}}(dt + W_{BD}d\phi)^2 + e^{2(k_{BD}-U_{BD})}[(dx^1)^2 + (dx^2)^2] +
h^2_{BD}e^{-2U_{BD}}d\phi^2
\end{eqnarray}
where $U$, $W$, $k$ and $h$ are functions of $x^1$ and $x^2$ only. The significance of the
choice of the metric in this form has been thoroughly discussed by Matzner and Misner [4]
and Misra and Pandey [5]. Tiwari and Nayak, and Singh and Rai first wrote down the Einstein-Maxwell
and BD-Maxwell field equations for the choice of metrics in eq.(3) and (4) respectively.
Comparing the two sets of field equations closely, they realized that stationary, axisymmetric
solutions of the BD-Maxwell field equations are obtainable from those of Einstein-Maxwell
field equations provided certain relations between metric functions hold. \\
That is, if $(W_{E}, ~k_{E}, ~U_{E}, ~h_{E}, ~A^{E}_{\mu})$ form a stationary, axisymmetric solution
to the Einstein-Maxwell field equations for the metric in eq.(3), then a corresponding
stationary, axisymmetric solution to the BD-Maxwell field equations for the metric in eq.(4) is
given by $(W_{BD}, ~k_{BD}, ~U_{BD}, ~h_{BD}, ~A^{BD}_{\mu})$ where
\begin{eqnarray}
W_{BD} &=& W_{E}, ~~~k_{BD} = k_{E}, ~~~U_{BD} = U_{E} - {1\over 2}\log \Phi, \\
h_{BD} &=& [h_{E}]^{(2\omega - 1)/(2\omega + 3)}, ~~~\Phi = [h_{E}]^{4/(2\omega + 3)},
~~~A^{BD}_{\mu} = A^{E}_{\mu}. \nonumber
\end{eqnarray}
Now what remains is to apply this method to obtain the Kerr-Newman-type solution in BD-Maxwell theory
(BDKN solution) from the known Kerr-Newman (KN) solution [6] in Einstein-Maxwell theory. And to do so, one
needs some preparation which involves casting the KN solution given in Boyer-Lindquist coordinates [7]
$(t, r, \theta, \phi)$ in the metric form in eq.(3) by performing a coordinate transformation (of
$r$ alone) suggested by Misra and Pandey [5]. Namely, we start with the KN solution written in 
Boyer-Lindquist coordinates
\begin{eqnarray}
ds^2 &=& -dt^2 + \Sigma (d\theta^2 + {dr^2\over \Delta}) + (r^2 + a^2)\sin^2 \theta d\phi^2 
+ {(2Mr - e^2)\over \Sigma}[dt - a\sin^2 \theta d\phi]^2, \nonumber \\
A_{\mu} &=& -{er\over \Sigma}[\delta^{t}_{\mu} - a\sin^2 \theta \delta^{\phi}_{\mu}] 
\end{eqnarray}
where $\Sigma = r^2+a^2 \cos^2 \theta $ and $\Delta = r^2 -2Mr + a^2 + e^2$ with $M$, $a$, and $e$
denoting the ADM mass, angular momentum per unit mass, and the electric charge respectively.
Consider now the transformation of the radial coordinate introduced by Misra and Pandey [5]
\begin{eqnarray}
r = e^{R} + M + {(M^2-a^2-e^2)\over 4}e^{-R}
\end{eqnarray}
which gives $dr^2/ \Delta = dR^2$. Then the KN solution can now be cast in the form in eq.(3),
i.e.,
\begin{eqnarray}
ds^2 = -[{\Delta - a^2\sin^2 \theta \over \Sigma}](dt + W d\phi)^2 + \Sigma [d\theta^2 + dR^2 +
{\Delta \sin^2 \theta \over {\Delta - a^2\sin^2 \theta}}d\phi^2]
\end{eqnarray}
with now $\Sigma = L^2+a^2 \cos^2 \theta $ and $\Delta = L^2 -2ML + a^2 + e^2$ where we set, as a
short-hand notation, $L \equiv e^{R} + M + {(M^2-a^2-e^2)\over 4}e^{-R}$.
Now we can read off the metric components as
\begin{eqnarray}
e^{2U_{E}} &=& [{\Delta - a^2\sin^2 \theta \over \Sigma}],
~~~W_{E} = {{a\sin^2 \theta (L^2 + a^2 - \Delta)}\over {\Delta - a^2\sin^2 \theta}}, \nonumber \\
e^{2k_{E}} &=& (\Delta - a^2\sin^2 \theta),
~~~h^2_{E} = \Delta \sin^2 \theta. 
\end{eqnarray}
Then using the rule in eq.(5) in the algorithm by Tiwari and Nayak, and Singh and Rai, we can now
construct BDKN solution in BD-Maxwell theory as
\begin{eqnarray}
ds^2 &=& - \left({{L^2 + a^2\cos^2 \theta - 2ML + e^2}\over {L^2 + a^2\cos^2 \theta}}\right)
(L^2 + a^2 - 2ML + e^2)^{-2/(2\omega+3)}\sin^{-4/(2\omega+3)}\theta \nonumber \\
&\times &\left[dt + {{a\sin^2 \theta (2ML-e^2)}\over {L^2 + a^2\cos^2 \theta - 2ML + e^2}}d\phi \right]^2 
\nonumber \\
&+& (L^2 + a^2 - 2ML + e^2)^{2/(2\omega+3)}\sin^{4/(2\omega+3)}\theta (L^2 + a^2\cos^2 \theta)
[d\theta^2 + dR^2] \nonumber \\
&+& (L^2 + a^2 - 2ML + e^2)^{(2\omega+1)/(2\omega+3)}\sin^{2(2\omega+1)/(2\omega+3)}\theta
\left({{L^2 + a^2\cos^2 \theta}\over {L^2 + a^2\cos^2 \theta - 2ML + e^2}}\right)d\phi^2, \nonumber \\
\Phi (R, \theta) &=& (L^2 + a^2 - 2ML + e^2)^{2/(2\omega+3)}\sin^{4/(2\omega+3)}\theta, \\
A_{\mu} &=& -{eL \over {L^2 + a^2\cos^2 \theta}}[\delta^{t}_{\mu} - a\sin^2 \theta \delta^{\phi}_{\mu}].
\nonumber
\end{eqnarray}
Then by transforming back to the standard Boyer-Lindquist coordinates using eq.(7), we finally arrive at
the BDKN solution in Boyer-Lindquist coordinates given by
\begin{eqnarray}
ds^2 &=& \Delta^{-2/(2\omega+3)}\sin^{-4/(2\omega+3)}\theta \left[ - \left({\Delta - a^2\sin^2 \theta \over
\Sigma}\right)dt^2 - {2a\sin^2 \theta (r^2+a^2-\Delta)\over \Sigma}dt d\phi \right. \nonumber \\
&+&\left. \left({(r^2+a^2)^2 - \Delta a^2\sin^2 \theta \over \Sigma }\right)\sin^2 \theta d\phi^2 \right]
+ \Delta^{2/(2\omega+3)}\sin^{4/(2\omega+3)}\theta \left[ {\Sigma \over \Delta}dr^2 + \Sigma d\theta^2 \right], 
\nonumber \\
\Phi (r, \theta) &=& \Delta^{2/(2\omega+3)}\sin^{4/(2\omega+3)}\theta,
~~~A_{\mu} = -{er\over \Sigma}[\delta^{t}_{\mu} - a\sin^2 \theta \delta^{\phi}_{\mu}]. 
\end{eqnarray}
Note that as $\omega \rightarrow \infty$, this BDKN solution goes over to the standard KN solution as
it should since the $\omega \rightarrow \infty$ limit of BD theory is the Einstein gravity.
Now that we have an exact, electrovac, stationary axisymmetric solution to the BD-Maxwell theory.
Then it is natural to ask if this BDKN solution can describe a black hole spacetime resulted from a
gravitational collapse. And if it can, furthermore one might be curious whether this BDKN solution
or its three special cases (i.e., BD-Schwarzschild, BD-Reissner-Nordstrom or BD-Kerr solutions) could
describe non-trivial black hole spacetimes which are {\it different} from those described by their
general relativistic counterparts. Indeed, questions of this sort had been raised long ago, and actually
Penrose [8] {\it conjectured} that even in BD theory of gravity, the relativistic gravitational collapse in three
spatial dimensions would produce black holes identical to those in general relativity. And this conjecture
received some support from the work of Thorne and Dykla [8] in which they presented four pieces of evidence
in favor of the conjecture by employing mainly the ``large-$\omega $'' expansion scheme (recall that in the
limit $\omega \rightarrow \infty$, the BD theory goes over to the general relativity). As Thorne and Dykla
mentioned in their work, however, the conjecture of Penrose was not fully proved since detailed analysis
of the collapse with arbitrary, finite values of the generic BD parameter $\omega$ is needed. In this regard,
we now seem to be in a better shape toward the serious investigation on the validity of the conjecture
since we have an exact, stationary axisymmetric solution possessing arbitrary $\omega $ values which was
not available at the time. Therefore we begin with the bottomline qualification for the BDKN solution
in BD-Maxwell theory to describe a rotating, charged black hole spacetime, namely the possible occurrence
of non-singular event horizon. Then along this line, perhaps the most natural first step is to ask under
what circumstances the Killing horizons develop. Just like the KN solution in Einstein-Maxwell theory,
this BDKN solution is stationary and axisymmetric and hence possesses the time translational Killing field
$\xi^{\mu}=(\partial /\partial t)^{\mu}$ and the rotational Killing field 
$\psi^{\mu}=(\partial /\partial \phi)^{\mu}$ correspondingly and it is their linear combination,
$\chi^{\mu} = \xi^{\mu} + \Omega_{H}\psi^{\mu}$ which is normal to the Killing horizons, if any
(here $\Omega_{H}$ denotes the angular velocity of the Killing horizon). And if Killing horizons are
present, they occur at points where $\chi^{\mu}$ becomes null which turn out to be zeroes of
$\Delta^{(2\omega+1)/(2\omega+3)}=0$. Thus first for $\omega = -1/2$, obviously no horizon occurs.
Next for $\left({2\omega +1 \over 2\omega +3}\right) > 0$, i.e., for $\omega < -3/2$ or $\omega > -1/2$,
two Killing horizons occur at $r_{\pm} = M \pm (M^2-a^2-e^2)^{1/2}$ (provided $M^2\geq a^2+e^2$) which
are precisely the same locations as those of Killing horizons of KN black holes in Einstein-Maxwell theory.
Now since the formation of horizons appears to be possible, next we investigate their nature. And to this
end, we examine behaviors of the invariant curvature polynomials such as  $R$, $R_{\mu\nu}R^{\mu\nu}$ and 
$R_{\mu\nu\alpha\beta}R^{\mu\nu\alpha\beta}$, the surface gravity $\kappa$ and the energy density of 
the BD scalar field $T^{BD}_{\mu \nu}\xi^{\mu}\xi^{\nu}$ on these candidates for Killing horizons. 
And as we mentioned above, since the bottomline qualification for the black hole interpretation of
BDKN solution is the regularity of the horizon candidate, we begin with the examination of behavior
of invariant curvature polynomials on the horizon candidate at which $\Delta = 0$.
First, the curvature scalar is calculated to be
\begin{eqnarray}
R &=& {\omega \over \Phi^2}g^{\alpha \beta}\nabla_{\alpha}\Phi \nabla_{\beta}\Phi +
{3\over \Phi}\nabla_{\alpha}\nabla^{\alpha}\Phi \\
&=& \omega \left({4\over 2\omega+3}\right)^2 {1\over \Sigma}\sin^{-4/(2\omega+3)}\theta 
[(r-M)^2\Delta^{-(2\omega+5)/(2\omega+3)} + \cot^2 \theta \Delta^{-2/(2\omega+3)}]. \nonumber
\end{eqnarray}
As was the case with KN black hole solutions, it also blows up at $\Sigma = 0$ (i.e., $r=0$,
$\theta = \pi/2$) indicating that the BDKN black hole solution also has the curvature singularity
with the same ``ring'' structure. The direct computation of the other two curvature polynomials,
i.e., the Ricci square $R_{\mu\nu}R^{\mu\nu}$ and the Kretschmann curvature invariant
$R_{\mu\nu\alpha\beta}R^{\mu\nu\alpha\beta}$ for this BDKN solution is a formidable job.
But a close inspection reveals that indeed we can save considerable amount of labor. Namely,
consider now the Brans-Dicke-Schwarzschild (BDS) spacetime solution that can be obtained by setting
$a = e = 0$ in the BDKN solution in eq.(11)
\begin{eqnarray}
ds^2 &=& \Delta^{-2/(2\omega+3)}\sin^{-4/(2\omega+3)}\theta
\left[-\left(1 - {2M\over r}\right)dt^2 + r^2 \sin^2 \theta d\phi^2 \right] \nonumber \\
&+& \Delta^{2/(2\omega+3)}\sin^{4/(2\omega+3)}\theta
\left[\left(1 - {2M\over r}\right)^{-1}dr^2 + r^2 d\theta^2 \right], \\
&\Phi& (r, \theta) = \Delta^{2/(2\omega+3)}\sin^{4/(2\omega+3)}\theta \nonumber
\end{eqnarray}
where now $\Delta = r(r - 2M)$. A remarkable feature of this BDS solution is the fact that,
unlike the Schwarzschild solution in general relativity, the spacetime it describes is 
static (i.e., non-rotating) but {\it not} spherically-symmetric. Thus computing
$R_{\mu\nu}R^{\mu\nu}$ and $R_{\mu\nu\alpha\beta}R^{\mu\nu\alpha\beta}$ for this BDS solution
and examining their behaviors on the horizon candidate at which $\Delta = 0$ would be sufficient
to envisage the possibility of regular event horizon for both BDS and BDKN solutions.
In addition, another noticeable characteristic of both BDKN and BDS spacetime solutions is that
they have possible coordinate singularities not only at the outer event horizon where $\Delta = 0$
but also along the symmetry axis $\theta = 0, ~\pi$. Thus in order to explore the nature of this
metric singularity along the symmetry axis, the computation of invariant curvature polynomials
looks necessary. The result of the computation of the Ricci square and
the Kretschmann curvature invariant for this BDS solution is given in the appendix.
And it is a straightforward matter to realize that the two invariant curvature polynomials
$R_{\mu\nu}R^{\mu\nu}$ and $R_{\mu\nu\alpha\beta}R^{\mu\nu\alpha\beta}$ given in the appendix and
the curvature scalar $R$ given above in eq.(12) become finite both on the horizon candidate at
which $\Delta = 0$ and along the symmetry axis $\theta = 0, ~\pi$ provided the generic BD
$\omega$-parameter takes values in the range $-5/2 \leq \omega < -3/2$.
Next, we turn to the computation of surface gravity at the Killing
horizons, $\kappa_{\pm}$. Generally, the surface gravity $\kappa$ is defined in a gravity theory-
independent manner as follows. Since the horizon is a null surface, there we have $\chi^{\mu}\chi_{\mu}=0$
where $\chi^{\mu}$ is the Killing field normal to the horizon we defined above. This implies that
$\nabla^{\mu}(\chi^{\nu}\chi_{\nu})$ is also normal to the horizon. Thus on the horizon, there exists a
function $\kappa$ such that
\begin{eqnarray}
\nabla^{\mu}(\chi^{\nu}\chi_{\nu}) = - 2\kappa \chi^{\mu} ~~~~{\rm or}
~~~~\chi^{\nu}\nabla_{\nu}\chi_{\mu} = \kappa \chi_{\mu}  \nonumber
\end{eqnarray}
from which it can be derived that
\begin{eqnarray}
\kappa^2 = -{1\over 2}(\nabla^{\mu}\chi^{\nu})(\nabla_{\mu}\chi_{\nu})
\end{eqnarray}
where the evaluation on the horizon is understood. Now, for the non-trivial BDKN black hole solution
at hand, a straightforward albeit somewhat tedious calculation yields
\begin{eqnarray}
\kappa_{\pm} = \Delta^{-2/(2\omega+3)}(r_{\pm})\sin^{-4/(2\omega+3)}\theta
\left({2\omega +1 \over 2\omega +3}\right) {(r_{\pm} - r_{\mp})\over 2(r^2_{\pm}+a^2)}.
\end{eqnarray}
Lastly, the energy density of the BD scalar field is computed using eq.(2) as
\begin{eqnarray}
&T^{BD}_{\mu \nu}&\xi^{\mu}\xi^{\nu} = T^{BD}_{tt} \\
&=& {1\over 8\pi}\Delta^{-4/(2\omega+3)}\sin^{-8/(2\omega+3)}\theta \left[{\omega \over 2}
\left({4\over 2\omega +3}\right)^2 \left({(r-M)^2\over \Delta} + \cot^2 \theta \right)
\left({\Delta - a^2\sin^2 \theta \over \Sigma^2}\right) \right. \nonumber \\
&+& \left({4\over 2\omega+3}\right){1\over \Sigma^2}\left\{\cot^2 \theta \left(\left({2\omega+1 \over
2\omega +3}\right)a^2\sin^2 \theta + \left({2\over 2\omega +3}\right)\Delta - a^2\sin^2 \theta
\left({\Delta - a^2\sin^2 \theta \over \Sigma}\right)\right) \right. \nonumber \\
&-& \left. \left. \left({r-M \over \Delta}\right)\left(\left({2\omega+1 \over 2\omega +3}\right)(r-M)\Delta +
\left({2\over 2\omega +3}\right)(r-M)a^2\sin^2 \theta - r\Delta 
\left({\Delta - a^2\sin^2 \theta \over \Sigma}\right)\right)\right\} \right]. \nonumber
\end{eqnarray}
It is interesting to note that this energy density of the BD scalar field also blows up at
the curvature singularity $\Sigma = 0$. \\
Now (i) for $\omega \rightarrow \infty$, $R=0$, $\kappa_{\pm}=(r_{\pm} - r_{\mp})/2(r^2_{\pm}+a^2)$
and $T^{BD}_{\mu \nu}\xi^{\mu}\xi^{\nu}=0$ on the surfaces $r=r_{\pm}$. 
This is an anticipated result since this is the correct KN black hole limit in Einstein-Maxwell theory. 
(ii) Next for $\infty > \omega > -1/2$, on the surfaces $r=r_{\pm}$, $(R, ~R_{\mu\nu}R^{\mu\nu}, 
~R_{\mu\nu\alpha\beta}R^{\mu\nu\alpha\beta}) \rightarrow \infty$, $\kappa_{\pm} \rightarrow \infty$ and
$T^{BD}_{\mu \nu}\xi^{\mu}\xi^{\nu}\rightarrow \infty$ with $\Phi (r_{\pm}, \theta)=0$. This
indicates that the surfaces $r=r_{\pm}$ are singular and fail to act as horizons and hence the
corresponding metric cannot describe a black hole spacetime. (iii) Finally for $\omega < -3/2$,
or more precisely for $-5/2 \leq \omega <-3/2$, on the surfaces $r=r_{\pm}$, $(R, ~R_{\mu\nu}R^{\mu\nu},
~R_{\mu\nu\alpha\beta}R^{\mu\nu\alpha\beta}) = 0$ (or $const.$
particularly for $\omega = -5/2$), $\kappa_{\pm}=0$ and $T^{BD}_{\mu \nu}\xi^{\mu}\xi^{\nu}=0$ with
$\Phi (r_{\pm}, \theta)\rightarrow \infty$. Namely the curvature invariants are finite, surface gravity 
is zero and the BD scalar field satisfies the weak energy condition although its value diverges there. 
(Here, infinite value of $\Phi$ indicates that the effective Newton's constant tends to zero.) Thus in this range
of the $\omega $-values, the surfaces $r=r_{\pm}$ may act as regular Killing horizons and hence
the corresponding BDKN metric solution appears to describe non-trivial black hole spacetimes
{\it different} from those in Einstein-Maxwell theory. In particular for $\omega = -5/2$, the
corresponding non-trivial BDKN black hole solution singles out with a relatively simple form
given by
\begin{eqnarray}
ds^2 &=& - \left[{\Delta - a^2\sin^2 \theta \over \Sigma}\right]\Delta \sin^2 \theta dt^2 
- {2a\sin^4 \theta (r^2+a^2-\Delta)\Delta \over \Sigma}dt d\phi \nonumber \\
&+& \left[{(r^2+a^2)^2 - \Delta a^2\sin^2 \theta \over \Sigma }\right]\Delta \sin^4 \theta d\phi^2 
+ {\Sigma \over \Delta^2 \sin^2 \theta }dr^2 + {\Sigma \over \Delta \sin^2 \theta } d\theta^2, \nonumber \\
\Phi (r, \theta) &=& {1\over \Delta \sin^2 \theta}.
\end{eqnarray}
Lastly, for the rest of the $\omega $-values, i.e., for $\omega <-5/2$, the surfaces $r=r_{\pm}$ are
singular horizons on which $(R,~R_{\mu\nu}R^{\mu\nu}, ~R_{\mu\nu\alpha\beta}R^{\mu\nu\alpha\beta})
\rightarrow \infty$, $\kappa_{+}=0$ and  $T^{BD}_{\mu \nu}\xi^{\mu}\xi^{\nu}=0$ and for
$-3/2<\omega \leq -1/2$, no Killing horizon develops and thus the curvature singularity at $\Sigma =0$
is naked. Therefore it now appears that for the values of the generic BD $\omega $-parameter in the
limited range $-5/2 \leq \omega < -3/2$, the BDKN solution in BD-Maxwell theory may describe non-trivial
black hole spacetimes. 
\\
{\bf III. Nature of BDKN black holes}
\\
Now that we have non-trivial BDKN black hole solutions. It seems then natural to explore its thermodynamics and 
causal structure in some more detail. Firstly, these BDKN black hole solutions have vanishing surface
gravity at the event horizon, $\kappa_{+}=0$ and hence {\it zero} Hawking temperature, 
$T_{H}=\kappa_{+}/2\pi=0$. In other words, they do not 
radiate and hence are completely ``dark and cold''. Certainly, this is a very bizzare feature in
sharp contrast to evaporating black holes in general relativity. Next, we turn to their causal structure.
As noted earlier, the two Killing horizons, i.e., the outer event horizon and the inner Cauchy horizon
turn out to occur precisely at the same locations (i.e., same coordinate distances)
as those of KN black hole solutions in Einstein-Maxwell
theory, i.e., at $r_{\pm} = M \pm (M^2-a^2-e^2)^{1/2}$. Also it is interesting to note that the proper
area of the event horizon at $r = r_{+}$,
\begin{eqnarray}
A = \int_{r_{+}} d\theta d\phi (g_{\theta \theta}g_{\phi \phi})^{1/2} = 4\pi (r^2_{+}+a^2)
\end{eqnarray}
is again exactly the same as that of standard KN black hole spacetime. In addition, its angular velocity 
at the event horizon coincides with that of standard KN solution as well 
\begin{eqnarray}
-W^{-1}_{BD}(r_{+}) = {a\over {r^2_{+}+a^2}} = -W^{-1}_{E}(r_{+}).
\end{eqnarray}
Next, observe that the norm of the time translational Killing field
\begin{eqnarray}
\xi^{\mu}\xi_{\mu} = g_{tt} = -\Delta^{-2/(2\omega+3)}\sin^{-4/(2\omega+3)}\theta
\left[{\Delta - a^2\sin^2 \theta \over \Sigma}\right]
\end{eqnarray}
goes like negative $(r_{-}<r<r_{+})$ $\rightarrow$ positive $(r_{+}<r<r_{s})$ $\rightarrow$ negative
$(r>r_{s})$ with $r_{s} = M + (M^2-a^2\cos^2 \theta - e^2)^{1/2} > r_{+}$ being the larger root of
$\xi^{\mu}\xi_{\mu}$, indicating that $\xi^{\mu}$ behaves as timelike $\rightarrow$ spacelike
$\rightarrow$ timelike correspondingly. And particularly the region in which $\xi^{\mu}$ stays
spacelike extends outside hole's event horizon. This region is the so-called ``ergoregion'' and
its outer boundary on which $\xi^{\mu}$ becomes null, i.e., $r=r_{s}$ is called ``static limit''
since inside of which no observer can possibly remain static. Thus if we recall the location of the
static limit in standard KN black hole solution, we can realize that even the locations of
ergoregions in two black hole spacetimes, KN and BDKN, are the same as well. Namely in two theories,
i.e., the BD-Maxwell theory and the Einstein-Maxwell theory, rotating, charged black hole solutions
turn out to possess {\it identical} causal structure (i.e., the locations of ring singularities,
two Killing horizons and static limits are the same) and hence exhibit the same global topology.
Thus actually what distinguishes the BDKN black hole spacetime from its general relativity's
counterpart, i.e., the KN black hole is the local geometry alone such as the curvature characterized by
the specific $\omega $-values, $-5/2 \leq \omega < -3/2$. At this point, perhaps it is relevant
to mention the behavior of the BD scalar field which plays unique role only in BD theory of gravity.
Independently of Penrose [8] and of Thorne and Dykla [8], Hawking [9] also explored the possible
existence of black hole solutions in BD theory and put forward a theorem which states that 
stationary black holes in BD theory are identical to those in general relativity. To be a little
more concrete, Hawking extended some of his theorems for general relativistic black holes to BD
theory and showed that any object collapsing to a black hole in BD theory must settle into
final equilibrium state which is either Schwarzschild or Kerr spacetime. And in doing so, he
``assumed'' that the BD scalar field $\Phi$ satisfies the weak energy condition and is constant
outside the black hole. Therefore now one may be puzzled as we realized in the present work that
non-trivial BDKN black hole solutions {\it different} from general relativistic KN solution
could exist in seemingly contradiction to Hawking's theorem. There is, however, no contradiction.
Hawking deduced the theorem by manipulating the BD field equations and most crucially ``assuming''
the strict conditions on the BD scalar field stated above but not by working with an explicit spacetime
solution which was not available at the time. In the present work, however, we investigated closely
the known, explicit stationary axisymmetric solution in BD theory. And in particular, when the 
BDKN solutions can describe non-trivial black hole spacetimes for the BD $\omega $-parametr
values $-5/2 \leq \omega < -3/2$, the accompanying BD scalar field solution 
$\Phi (r, \theta) = \Delta^{2/(2\omega+3)}\sin^{4/(2\omega+3)} \theta$ turns out {\it not} to
be a constant field outside the event horizon at $r=r_{+}$. Besides, the energy density of this
BD scalar field, $T^{BD}_{\mu \nu}\xi^{\mu}\xi^{\nu}$ whose explicit form was given earlier in eq.(16)
does not strictly obey the weak energy condition for all $r$. Namely the value of
$T^{BD}_{\mu \nu}\xi^{\mu}\xi^{\nu}$ does not remain non-negative for all $r$. Rather, its
value and hence the signature changes from point to point. In short, the Hawking's theorem simply
cannot be applied to the present situation and hence the results of the present study needs not
be restricted by Hawking's theorem. At this point, it seems appropriate to ask whether the non-trivial
BDKN black hole solution studied in the present work can be viewed as a counterexample to the
no-hair theorem of black holes. In the loose sense, one may think of the non-trivial behavior
of the BD scalar field outside the event horizon as indicating the appearance of ``scalar hair''.
Here, however, we need to be more precise with the nature of no-hair theorem. Following Bizon [11],
for instance, a certain theory is said to allow a hairy black hole solution if there is a need to
specify quantities other than conserved charges defined at asymptotic infinity such as the mass,
angular momentum and the electric charge in order to characterize comletely a stationary black hole
solution within that theory. Thus in this stricter sense, the non-trivial BDKN black hole solution
studied in the present work does not constitute a hairy black hole solution since both the metric
and BD scalar field solutions in eq.(11) are specified completely by the ADM mass $M$, angular
momentum per unit mass $a$, and the electric charge $e$ only and no other quantities.
\\
{\bf IV. Discussions}
\\
Before we address the physical implication of the non-trivial BDKN black hole solution found in 
the present work, we would like to comment on a technical issue, i.e., the divergent behavior
of the BD scalar field solution on the horizon. And in relation to this, it is interesting to note 
that our BDKN black hole solution shares two peculiar features, i.e., the divergent behavior of the
scalar field on the horizon and the null Hawking radiation, with the well-known Bekenstein black
hole solution in Einstein-conformal scalar field theory [12]. Namely, using a suitable solution
generation technique, long ago, Bekenstein constructed a static, spherically-symmetric black hole
solution in which the metric part corresponds to that of extreme Reissner-Nordstrom (RN) black hole
and thus represents non-radiating black hole spacetime and the conformal scalar field solution
diverges on the horizon. Therefore for direct and parallel comparison between the two solutions,
it seems appropriate to take the BDS solution analyzed in detail in the appendix. The two theories,
of course, have completely different nature and motivations. The Einstein-conformal scalar field
theory has been devised guided mainly by a particular (Weyl) symmetry and the scalar field there is
supposed to describe a matter. The BD theory, on the other hand, is an alternative theory to Einstein
gravity and the BD scalar field here represents a spacetime-varying effective Newton's constant, not
a matter. Thus the divergent behavior of the scalar field in Einstein-conformal scalar field theory
could be disastrous but that of the BD scalar field in BD theory essentially represents the
vanishing effective Newton's constant in a certain region of spacetime. Besides, since the energy
density of the BD scalar field $T^{BD}_{\mu\nu}\xi^{\mu}\xi^{\nu}$ given in eq.(16) vanishes and
hence satisfies the weak energy condition on the horizon at which $\Delta = 0$ (of course for
$-5/2 \leq \omega <-3/2$), we do not worry too much about the divergent behavior of the BD scalar
field there.  Another interesting contrast is that the metric solution of
Bekenstein black hole spacetime there corresponds to a familiar extreme RN metric which is static and 
spherically-symmetric whereas the BDS metric solution here exhibits a remarkable feature that it is
static (i.e., non-rotating) but not spherically-symmetric as we pointed out earlier. Besides, the
Bekenstein solution represents a ``hairy'' black hole [13] whereas our BDS black hole solution
carries no hair as we mentioned above. Now, the point of central interest we would like to make 
is about the issue raised
recently by Sudarsky and Zannias [13]. To be a little more concrete, they showed that the
divergent behavior of the conformal scalar field solution on the horizon essentially leads the Bekenstein
black hole solution to fail to satisfy Einstein field equations particularly on the horizon. They, thus,
concluded that the Bekenstein solution cannot be considered as a genuine black hole solution and
therefore the black hole no-hair theorem is saved. And as a manifest evidence for their argument against
the black hole interpretation, Sudarsky and Zannias demonstrated that by working in Eddington-Finkelstein
null coordinates, parts of the Einstein field equations can be shown not to hold
in a rigorous sense as the left-hand side of the equation, say $R_{\mu\nu}$ vanishes on the
horizon (since the metric is that of extreme RN) while the right-hand side, namely the energy-momentum
tensor of the conformal scalar field, $8\pi [T_{\mu\nu}-g_{\mu\nu}T^{\lambda}_{\lambda}/2]$ is 
ill-defined on the horizon as the conformal scalar field diverges there. Thus with this in mind,
now we consider the validity of our BDS solution on the horizon. 
As one can see in the appendix, the 5 non-vanishing components
of BD field equations, i.e., $tt$, $rr$, $r\theta$, $\theta \theta$ and $\phi \phi$ parts appear
to hold perfectly. In particular, on the horizon and for $-5/2 \leq \omega < -3/2$, $tt$ and $\phi \phi$
parts hold as ``$0 = 0$''. The other 3 parts, however, hold as ``$\infty = \infty$''. Namely, in these 3
equations, not only the energy-momentum tensor of the BD scalar field on the right-hand side of BD
field equations but also the Ricci tensor components on the left-hand side diverges in exactly
the same manner on the horizon. Indeed precisely these 3 equations are the ones we need to be
careful with. Here, however, we must say that a naive attempt toward the validity check of the
BDS solution on the horizon in exactly the same way as Sudarsky and Zannias did for Bekenstein
solution seems to be obscured. In fact,
the demonstration of Sudarsky and Zannias was successful largely because
the simple transformation from spherical to Eddington-Finkelstein null coordinates for the Bekenstein
solution was available. Indeed, the virtue of null coordinates is that in terms of which
$g_{rr}=0$ and hence $l^{\mu} = (\partial/\partial r)^{\mu}$ becomes a smooth null vector field
such that the quantity $R_{\mu\nu}l^{\mu}l^{\nu}$ can be shown to be finite (zero) on the horizon
while the right-hand side of Einstein equation, $8\pi [T_{\mu\nu}-g_{\mu\nu}T^{\lambda}_{\lambda}/2]
l^{\mu}l^{\nu}$ is ill-defined. In contrast, however, taking a Eddington-Finkelstein-type null
coordinates for the BDS solution is not so obvious as in the case of static, spherically-symmetric
black hole solutions in Einstein theory and is indeed practically awkward.
It is essentially due to the peculiar feature of BDS metric solution
which is static but not spherically-symmetric as has been stressed earlier. Thus in the present
work, we do not pursue validity check of the BDS solution on the horizon in this direction
and leave it as an issue for future investigation. As a result, for
the 3 parts of the BD field equations yielding ``$\infty = \infty$'', no definite statement
can be made yet concerning whether or not they are valid, i.e., these 3 equations are really 
satisfied on the horizon. Nevertheless, one interesting point we can make is that if the BDS
solution does represent a genuine black hole spacetime, then the black hole no-hair theorem
appears to survive even in the BD theory of gravity since the BDS (and BDKN as well) spacetime
is not a hairy black hole solution as we discussed earlier. \\
It seems that now the most relevant question to ask is ; if they are genuine, would these
non-trivial BDKN black hole spacetimes in BD theory  of gravity exhibiting bizzare features such
as null radiation really arise in nature? Of course
this question needs to be answered very carefully and honestly and the answer for now does not seem
to be in the affirmative. Firstly, from field theory's viewpoint, the generic BD theory $\omega $-parameter
has to be ``positive'' in order for the BD scalar field $\Phi$ to have canonical (positive-definite)
kinetic energy as can be seen in the BD gravity theory action given in eq.(1). Secondly, it is well-known
that the BD gravity theory is in reasonable accord with all available observations and experiments
thus far provided $|\omega | \gtrsim 500$ [10]. Since both these constraints on the values of the 
$\omega $-parameter seem to rule out the range $-5/2 \leq \omega < -3/2$ in which the BDKN solution
could describe non-trivial black hole spacetimes, for now it seems fair to say that these non-trivial 
BDKN black hole spacetimes different from their general relativistic counterparts are unlikely to arise
in nature. This, however, may not be the end of the story. As we have seen in this work, the energy density 
of the explicit BD scalar field solution (which essentially consists of its kinetic energy) turns out not 
to satisfy the weak energy condition irrespective of the $\omega $-value.
Perhaps this implies that we may abandon the ``canonical kinetic 
energy'' condition on the BD scalar field and allow negative-$\omega $ values. Moreover, the lower bound 
$|\omega | \gtrsim 500$ may be relaxed considerably with the advances in technology associated with 
astronomical observations and astrophysical experiments. Thus perhaps it might be wise to keep the
possibility of non-trivial BDKN black holes alive. 
As a matter of fact, there is another type of possibility of greater physical significance and relevance. Note 
that the generic BD $\omega $-parameter is a kind of coupling constant appearing in the BD gravity action.
Thus in principle, it should be considered as a ``running'' coupling constant as a result of renormalization
in the quantum gravity context. And its scale-dependent behavior can be envisaged as follows. In the BD
gravity action given in eq.(1), the term $\sim \omega (\nabla_{\alpha}\Phi \nabla^{\alpha}\Phi /\Phi)$,
like other terms in the action, should be finite. Thus large-$\omega$ indicates the regime where the BD
scalar field $\Phi$ remains nearly constant which corresponds to the large-scale present universe limit
(in which the BD theory goes over to the general relativity). On the other hand, small-$\omega$ indicates 
the regime where the BD scalar field varies sizably with space and time which would presumably correspond
to the small-scale early universe limit. Thus if we are willing to accept the BD theory as a ``better''
effective theory of quantum gravity than general relativity to describe the entire stages (scales) of
the universe evolution, then at early times when the value of $\omega$ was small such as
$-5/2 \leq \omega < -3/2$, the non-trivial BDKN black holes like the ones studied in this work would
have had a chance to form. These ``primordial'' black holes, unlike their general relativistic counterparts,
however, do not evaporate as we discussed earlier. Thus it can be speculated that they might still hide
somewhere in the dark side of the space today as a possible constituent of the cold dark matter.
After all, new discoveries can be made when we keep our minds as well as eyes open.

\vspace*{1cm}

\begin{center}
{\rm\bf Acknowledgements}
\end{center}

This work was supported by grant of Post-doc. Program at Kyungpook National University (1998).

\vspace*{2cm}

\begin{center}
{\rm\bf Appendix : Computation of invariant curvature polynomials}
\end{center}

In this appendix, we shall explicitly write down the BD field equations satisfied 
by the BDS solution given in eq.(13) and then provide the result of 
the computation of its invariant curvature polynomials such as the Ricci square
$R_{\mu\nu}R^{\mu\nu}$ and the Kretschmann curvature invariant
$R_{\mu\nu\alpha\beta}R^{\mu\nu\alpha\beta}$. \\
First we start with the BD field equations.
The tensor part of the vacuum BD field equations that this BDS solution satisfies,

\begin{eqnarray}
R_{\mu\nu} = {\omega \over \Phi^2}\nabla_{\mu}\Phi \nabla_{\nu}\Phi +
{1\over \Phi}\nabla_{\mu}\nabla_{\nu}\Phi
\end{eqnarray}
has the following 5 non-vanishing components. Here we show that actually they are all satisfied
by the BDS solution in the sense that the explicit computation of the left-hand side, i.e., $R_{\mu\nu}$
and the right-hand side, i.e., ${\omega \over \Phi^2}\nabla_{\mu}\Phi \nabla_{\nu}\Phi +
{1\over \Phi}\nabla_{\mu}\nabla_{\nu}\Phi $ performed with the BDS solution given in eq.(13)
precisely agree. 
\small
\begin{eqnarray}
R_{tt} &=& {\omega \over \Phi^2}\nabla_{t}\Phi \nabla_{t}\Phi +
{1\over \Phi}\nabla_{t}\nabla_{t}\Phi \nonumber \\
&=& {1\over r^4}\Delta^{-4/(2\omega+3)}\sin^{-8/(2\omega+3)}\theta
{4\over (2\omega+3)}\left[{2\over (2\omega+3)}(r-M)^2 - M(r-M) + {2\over (2\omega+3)}
\cot^2 \theta \Delta \right], \nonumber \\
R_{rr} &=& {\omega \over \Phi^2}\nabla_{r}\Phi \nabla_{r}\Phi +
{1\over \Phi}\nabla_{r}\nabla_{r}\Phi \nonumber \\
&=& {1\over \Delta^2} {4\over (2\omega+3)}\left[{-4\over (2\omega+3)}(r-M)^2 + M(r-M) 
+ \left\{1 + {2\over (2\omega+3)}\cot^2 \theta \right\}\Delta \right], \nonumber \\
R_{r\theta} &=& {\omega \over \Phi^2}\nabla_{r}\Phi \nabla_{\theta}\Phi +
{1\over \Phi}\nabla_{r}\nabla_{\theta}\Phi \nonumber \\
&=& {1\over \Delta}\cot \theta 
{4\over (2\omega+3)}\left[{4\omega \over (2\omega+3)}(r-M) - (r-2M)\right], \\
R_{\theta \theta} &=& {\omega \over \Phi^2}\nabla_{\theta}\Phi \nabla_{\theta}\Phi +
{1\over \Phi}\nabla_{\theta}\nabla_{\theta}\Phi \nonumber \\
&=& {1\over \Delta}{4\over (2\omega+3)}\left[{2\over (2\omega+3)}(r-M)^2 + (r-M)(r-2M) - \left\{1 -
\left({2\omega-1\over 2\omega+3}\right)\cot^2 \theta \right\}\Delta \right], \nonumber \\
R_{\phi \phi} &=& {\omega \over \Phi^2}\nabla_{\phi}\Phi \nabla_{\phi}\Phi +
{1\over \Phi}\nabla_{\phi}\nabla_{\phi}\Phi \nonumber \\
&=& {-1\over \Delta}\Delta^{-4/(2\omega+3)}\sin^{-8/(2\omega+3)}\theta \times \nonumber \\
&&{4\over (2\omega+3)}\left[\left\{{2\over (2\omega+3)}(r-M)^2 - (r-M)(r-2M)\right\}\sin^2 \theta -
\left({2\omega+1\over 2\omega+3}\right)\cos^2 \theta \Delta \right]. \nonumber
\end{eqnarray}
\normalsize
And the scalar part of the vacuum BD field equations can easily be seen to hold. \\
Next we turn to the expressions for the invariant curvature polynomials. The Ricci square is
calculated to be
\small
\begin{eqnarray}
&&R_{\mu\nu}R^{\mu\nu} = g^{\mu\alpha}g^{\nu\beta}R_{\mu\nu}R_{\alpha\beta}  \\
&=& {1\over r^4}\Delta^{-2(2\omega+5)/(2\omega+3)}\sin^{-8/(2\omega+3)}\theta
\left[\left\{{4\over (2\omega+3)}\left[{2\over (2\omega+3)}(r-M)^2 - M(r-M) \right. \right. \right. \nonumber \\ 
&+& \left. \left. {2\over (2\omega+3)}\cot^2 \theta \Delta \right]\right\}^2  \nonumber \\
&+& \left\{{4\over (2\omega+3)}\left[{-4\over (2\omega+3)}(r-M)^2 + M(r-M) + \left(1 + {2\over (2\omega+3)}
\cot^2 \theta \right)\Delta \right]\right\}^2  \nonumber \\
&+& \left\{{4\over (2\omega+3)}\left[{2\over (2\omega+3)}(r-M)^2 + (r-M)(r-2M) - \left(1 - 
\left({2\omega-1\over 2\omega+3}\right)\cot^2 \theta \right)\Delta \right]\right\}^2  \nonumber \\
&+& \left. \left\{{4\over (2\omega+3)}\left[{2\over (2\omega+3)}(r-M)^2 - (r-M)(r-2M) - 
\left({2\omega+1\over 2\omega+3}\right)\cot^2 \theta \Delta \right]\right\}^2\right]  \nonumber \\
&+& {2\over r^4}\Delta^{-(2\omega+7)/(2\omega+3)}\sin^{-2(2\omega+7)/(2\omega+3)}\theta
\cos^2 \theta \left[\left\{{4\over (2\omega+3)}\left[{4\omega \over (2\omega+3)}(r-M) - (r-2M)\right]
\right\}^2\right]. \nonumber
\end{eqnarray}
\normalsize
It is worth noting that this Ricci square vanishes in the limit $\omega \rightarrow \infty$ as it should
since in which the BDS solution goes over to the Schwarzschild solution. 
And the Kretschmann curvature invariant is computed to be
\small
\begin{eqnarray}
&&R_{\mu\nu\alpha\beta}R^{\mu\nu\alpha\beta} = g^{\mu\sigma}g^{\nu\rho}g^{\alpha\lambda}g^{\beta\delta}
R^{\mu}_{\nu\alpha\beta}R^{\sigma}_{\rho\lambda\delta}  \\
&=& {2\over r^4}\Delta^{-2(2\omega+5)/(2\omega+3)}\sin^{-8/(2\omega+3)}\theta
\left[4\left\{2{(2\omega+5)\over (2\omega+3)^2}(r-M)^2 - \left({2\omega+7\over 2\omega+5}\right)M(r-M)
 + M^2  \right. \right. \nonumber \\
&-& \left. {1\over (2\omega+3)}\left[1 + {2\over (2\omega+3)}\cot^2 \theta\right]\Delta \right\}^2 
\nonumber \\
&+& 4\left\{\left[{2\over (2\omega+3)}(r-M) + (r-2M)\right]\left[{2\over (2\omega+3)}(r-M) - M \right]
- {1\over (2\omega+3)}\left[1 + {2\over (2\omega+3)}\cot^2 \theta\right]\Delta \right\}^2  \nonumber \\
&+& \left\{\left[{2\over (2\omega+3)}(r-M) - (r-2M)\right]\left[{2\over (2\omega+3)}(r-M) - M \right]
- 2{(2\omega+1)\over (2\omega+3)^2}\cot^2 \theta \Delta \right\}^2  \nonumber \\
&+& \left. \left\{\left({2\omega+7 \over 2\omega+3}\right)(r-M)\left[{2\over (2\omega+3)}(r-M) - (r-2M)\right]
+ \left({2\omega+1 \over 2\omega+3}\right)\left[1 + {2\over (2\omega+3)}\cot^2 \theta\right]\Delta \right\}^2 
\right] \nonumber \\
&+& {2\over r^4}\Delta^{-2(2\omega+5)/(2\omega+3)}\sin^{-4(2\omega+5)/(2\omega+3)}\theta
\left[\left\{\left[{2\over (2\omega+3)}(r-M) + (r-2M)\right] \times \right. \right. \nonumber \\
&&\left. \left[{2\over (2\omega+3)}(r-M) - M \right]
\sin^2 \theta - {2\over (2\omega+3)}\left[1 + {2\over (2\omega+3)}\cos^2 \theta\right]\Delta \right\}^2  
\nonumber \\
&+& \left\{\left[{2\over (2\omega+3)}(r-M) - (r-2M)\right]\left[{2\over (2\omega+3)}(r-M) - M \right]
\sin^2 \theta - 2{(2\omega+1)\over (2\omega+3)^2}\cos^2 \theta \Delta \right\}^2  \nonumber \\
&+& 2\left\{(r-M)\left[{2\over (2\omega+3)}(r-M) + (r-2M)\right]\sin^2 \theta + 
\left[{2\over (2\omega+3)} - \left({2\omega+5 \over 2\omega+3}\right)\sin^2 \theta \right]\Delta \right\}^2
\nonumber \\
&+& \left\{\left({2\omega+7 \over 2\omega+3}\right)(r-M)\left[{2\over (2\omega+3)}(r-M) - (r-2M)\right]
\sin^2 \theta + \left({2\omega+1 \over 2\omega+3}\right)\left[1 - \left({2\omega+5 \over 2\omega+3}\right)
\cos^2 \theta\right]\Delta \right\}^2 \nonumber \\
&+& \left\{\left[{2\over (2\omega+3)}(r-M) + (r-2M)\right]\left[{2\over (2\omega+3)}(r-M) - M \right]
\sin^2 \theta \right.  \nonumber \\
&-& \left. {2\over (2\omega+3)}\left[1 + {4\over (2\omega+3)}\cos^2 \theta\right]\Delta \right\}^2 \\
&+& \left. 2\left\{\left[{4\over (2\omega+3)^2}(r-M)^2 - (r-2M)^2\right]\sin^2 \theta +
\left({2\omega+1 \over 2\omega+3}\right)\left[1 - \left({2\omega-1 \over 2\omega+3}\right)
\cos^2 \theta\right]\Delta \right\}^2 \right] \nonumber \\
&+& {32\over r^4}\Delta^{-(2\omega+7)/(2\omega+3)}\sin^{-2(2\omega+7)/(2\omega+3)}\theta \cos^2 \theta
\left[\left\{{1\over (2\omega+3)}\left[{6\over (2\omega+3)}(r-M) + (r-4M)\right]\right\}^2 \right. \nonumber \\
&+& \left. \left\{{1\over (2\omega+3)}\left[{4\omega\over (2\omega+3)}(r-M) + (r-2M)\right]\right\}^2 \right].
\nonumber
\end{eqnarray}
\normalsize
Again, it is straightforward to check that this Kretschmann curvature invariant reduces to
that of Schwarzschild solution, $48M^2/r^6$ in the Einstein gravity limit 
(i.e., as $\omega \rightarrow \infty$) as it should.
In conclusion, a close inspection reveals that $R_{\mu\nu}R^{\mu\nu}$ and
$R_{\mu\nu\alpha\beta}R^{\mu\nu\alpha\beta}$ are finite (more precisely vanishes) both on the horizon 
candidate at which 
$\Delta = 0$ and along the symmetry axis $\theta = 0, ~\pi$ provided the generic BD $\omega $-parameter
takes values in the range $-5/2 \leq \omega <-3/2$.

\vspace*{2cm}

\noindent

\begin{center}
{\rm\bf References}
\end{center}

\begin{description}

\item {[1]} C. Brans and C. H. Dicke, Phys. Rev. {\bf 124}, 925 (1961).
\item {[2]} S. Weinberg, {\it Gravitation and Cosmology} (Wiley, New York, 1972).
\item {[3]} A. I. Janis, D. C. Robinson, and J. Winicour, Phys. Rev. {\bf 186}, 1729 (1969) ;
            H. A. Buchdahl, Int. J. Theor. Phys. {\bf 6}, 407 (1972) ;
            C. B. G. McIntosh, Commun. Math. Phys. {\bf 37}, 335 (1974) ;
            B. O. J. Tupper, Nuovo Cimento {\bf 19B}, 135 (1974) ;
            R. N. Tiwari and B. K. Nayak, Phys. Rev. {\bf D14}, 2502 (1976) ;
            J. Math. Phys. {\bf 18}, 289 (1977) ;
            T. Singh and L. N. Rai, Gen. Rel. Grav. {\bf 11}, 37 (1979).
\item {[4]} R. A. Matzner and C. W. Misner, Phys. Rev. {\bf 154}, 1229 (1967).
\item {[5]} R. M. Misra and D. B. Pandey, J. Math. Phys. {\bf 13}, 1538 (1972).
\item {[6]} R. P. Kerr, Phys. Rev. Lett. {\bf 11}, 237 (1963) ;
            E. J. Newman, E. Couch, K. Chinapared, A. Exton, A. Prakash, and R. Torrence,
            J. Math. Phys. {\bf 6}, 918 (1965).
\item {[7]} R. H. Boyer and R. W. Lindquist, J. Math. Phys. {\bf 8}, 265 (1967).
\item {[8]} R. Penrose, Lecture at Fifth Texas Symposium on Relativistic Astrophysics,
            Austin, Texas, December 16, 1970 ;
            K. S. Thorne and J. J. Dykla, Astrophys. J. {\bf L35}, 166 (1971).
\item {[9]} S. W. Hawking, Commun. Math. Phys. {\bf 25}, 167 (1972).
\item {[10]} See, for example, C. M. Will, {\it Theory and Experiment in Gravitational
             Physics}, revised edition (Cambridge Univ. Press, Cambridge 1993).
\item {[11]} P. Bizon, preprint, Jagellonian Univ. Instit. of Phys. Cracow Pol. (1994). 
\item {[12]} J. D. Bekenstein, Ann. Phys. {\bf 82}, 535 (1974) ; {\it ibid} {\bf 91},
             72 (1975).
\item {[13]} D. Sudarsky and T. Zannias, Phys. Rev. {\bf D58}, 087502 (1998) ; we thank the
             referee for drawing our attention to this reference.

\end{description}

\end{document}